\documentstyle[psfig]{mn}

\newif\ifAMStwofonts

\def\ee {1E~1841$-$045}
\def\ltsima{$\; \buildrel < \over \sim \;$}
\def\lsim{\lower.5ex\hbox{\ltsima}}
\def\gtsima{$\; \buildrel > \over \sim \;$}
\def\gsim{\lower.5ex\hbox{\gtsima}}
\def\approxlt{\mathrel{\spose{\lower 3pt\hbox{$\sim$}}
        \raise 2.0pt\hbox{$<$}}}
\def\approxgt{\mathrel{\spose{\lower 3pt\hbox{$\sim$}}
        \raise 2.0pt\hbox{$>$}}}
\newcommand{\be}{\begin{equation}}
\newcommand{\en}{\end{equation}}
\def\deg {^\circ}

\def\pdot {\dot P}

\def\msole{~M_{\odot}}

\def\aa #1 #2 {A\&A {#1} #2}
\def\aas #1 #2 {A\&AS {#1} #2}
\def\araa #1 #2 {ARA\&A {#1} #2}
\def\mnras #1 #2 {MNRAS {#1} #2}
\def\apj #1 #2 {ApJ {#1} #2}
\def\apjs #1 #2 {ApJS {#1} #2}
\def\apjl #1 #2 {ApJ {#1} #2}
\def\aj #1 #2 {AJ {#1} #2}
\def\nat #1 #2 {Nature {#1} #2}
\def\pasj #1 #2 {PASJ {#1} #2}
\def\pasp #1 #2 {PASP {#1} #2}

\ifoldfss
  \ifCUPmtlplainloaded \else
    \NewTextAlphabet{textbfit} {cmbxti10} {}
    \NewTextAlphabet{textbfss} {cmssbx10} {}
    \NewMathAlphabet{mathbfit} {cmbxti10} {} 
    \NewMathAlphabet{mathbfss} {cmssbx10} {} 
  \fi
  \ifAMStwofonts
    \ifCUPmtlplainloaded \else
      \NewSymbolFont{upmath} {eurm10}
      \NewSymbolFont{AMSa} {msam10}
      \NewMathSymbol{\upi}     {0}{upmath}{19}
      \NewMathSymbol{\umu}     {0}{upmath}{16}
      \NewMathSymbol{\upartial}{0}{upmath}{40}
      \NewMathSymbol{\leqslant}{3}{AMSa}{36}
      \NewMathSymbol{\geqslant}{3}{AMSa}{3E}

    \fi
  \fi
\fi 

\ifnfssone
  \newmathalphabet{\mathit}
  \addtoversion{normal}{\mathit}{cmr}{m}{it}
  \addtoversion{bold}{\mathit}{cmr}{bx}{it}
  \newmathalphabet{\mathbfit} 
  \addtoversion{normal}{\mathbfit}{cmr}{bx}{it}
  \addtoversion{bold}{\mathbfit}{cmr}{bx}{it}
  \newmathalphabet{\mathbfss} 
  \addtoversion{normal}{\mathbfss}{cmss}{bx}{n}
  \addtoversion{bold}{\mathbfss}{cmss}{bx}{n}
  \ifAMStwofonts
    \ifCUPmtlplainloaded \else
      \UseAMStwoboldmath
      \makeatletter
      \new@mathgroup\upmath@group
      \define@mathgroup\mv@normal\upmath@group{eur}{m}{n}
      \define@mathgroup\mv@bold\upmath@group{eur}{b}{n}
      \edef\UPM{\hexnumber\upmath@group}
      \new@mathgroup\amsa@group
      \define@mathgroup\mv@normal\amsa@group{msa}{m}{n}
      \define@mathgroup\mv@bold\amsa@group{msa}{m}{n}
      \edef\AMSa{\hexnumber\amsa@group}
      \makeatother
      \mathchardef\upi="0\UPM19
      \mathchardef\umu="0\UPM16
      \mathchardef\upartial="0\UPM40
      \mathchardef\leqslant="3\AMSa36
      \mathchardef\geqslant="3\AMSa3E
    \fi
  \fi
\fi 

\ifnfsstwo
  \DeclareMathAlphabet{\mathbfit}{OT1}{cmr}{bx}{it}
  \SetMathAlphabet\mathbfit{bold}{OT1}{cmr}{bx}{it}
  \DeclareMathAlphabet{\mathbfss}{OT1}{cmss}{bx}{n}
  \SetMathAlphabet\mathbfss{bold}{OT1}{cmss}{bx}{n}
  \ifAMStwofonts
    \ifCUPmtlplainloaded \else
      \DeclareSymbolFont{UPM}{U}{eur}{m}{n}
      \SetSymbolFont{UPM}{bold}{U}{eur}{b}{n}
      \DeclareSymbolFont{AMSa}{U}{msa}{m}{n}
      \DeclareMathSymbol{\upi}{0}{UPM}{"19}
      \DeclareMathSymbol{\umu}{0}{UPM}{"16}
      \DeclareMathSymbol{\upartial}{0}{UPM}{"40}
      \DeclareMathSymbol{\leqslant}{3}{AMSa}{"36}
      \DeclareMathSymbol{\geqslant}{3}{AMSa}{"3E}
    \fi
  \fi
\fi 

\ifCUPmtlplainloaded \else
  \ifAMStwofonts \else 
    \def\upi{\pi}
    \def\umu{\mu}
    \def\upartial{\partial}
  \fi
\fi

\title{A Search for the Optical/Infrared Counterpart of the Anomalous X-ray Pulsar \ee\thanks{Based on observations collected at 
the European Southern Observatory, La Silla, Chile. }
 }

\author[S. Mereghetti et al.]
  {S.~Mereghetti,$^1$
  R.P.~Mignani,$^2$
  S.~Covino,$^3$
  S.~Chaty,$^4$
  G.L.~Israel,$^{5,*}$
  R.~Neuh\"auser,${^6}$   
  \and H.~Plana,$^7$
  L.~Stella.$^{5,*}$\\
  $^1$Istituto di Fisica Cosmica del C.N.R., Via Bassini 15, I-20133 Milano,
Italy;  sandro@ifctr.mi.cnr.it \\
  $^2$E.S.O., Karl-Schwarzschild-Str. 2, D-85748 Garching bei Munchen, Germany \\
  $^3$Osservatorio Astronomico di Merate, v. E. Bianchi 46, I-22055 Merate (LC), Italy   \\
  $^4$Department of Physics and Astronomy,
  The Open University, Walton Hall, Milton Keynes,
  MK7 6AA, United Kingdom     \\
  $^5$Osservatorio Astronomico di Roma, Via Frascati 23,
I-00040 Monteporzio Catone (Roma), Italy; \\
  $^6$MPI Extraterrestrische Physik, D-85740 Garching, Germany  \\
  $^7$Observatorio Astron\'omico Nacional,
  Apartado Postal 877,22800 Ensenada, B.C., M\'exico \\
  $^*$Affiliated to ICRA.}

\date{Accepted 2000    .
      Received 2000    ;
      in original form      }

\pagerange{\pageref{firstpage}--\pageref{lastpage}}
\pubyear{2000}

\begin{document}

\maketitle

\label{firstpage}

\begin{abstract}
 
We have carried out a search for the optical and infrared counterpart of the Anomalous 
X--ray Pulsar \ee~, which is  located at the center of the supernova remnant Kes~73.
We present the first deep optical and infrared images of the field of \ee~, 
as well as  optical spectroscopy results that exclude the brightest objects in 
the error circle as possible counterparts. A few of the more reddened objects 
in this region can be considered as particularly interesting candidates, 
in consideration of the distance and absorption expected from the association with 
Kes~73. The strong interstellar absorption in the direction of the source does not allow 
to completely exclude the presence of main sequence massive companions.

\end{abstract}


\section{Introduction}

The X-ray source \ee~ was discovered  with the Einstein Observatory in 1979
(Kriss et al. 1985) at the center of the supernova remnant Kes~73 (G27.4+0.0), 
and studied in more detail with ROSAT (Helfand et al. 1994). Kes 73 is a young 
supernova remnant that shows a shell morphology in the radio band, without any
evidence for a central synchrotron nebula originating from a rotation-powered 
pulsar. The most likely interpretation for \ee, on the basis of 
these earlier observations, was that of an accretion powered system.

Recent data obtained with ASCA led to the discovery of a periodicity at 11.8 s and 
a period derivative of 4 10$^{-11}$ s s$^{-1}$ (Vasisht \& Gotthelf 1997). 
The corresponding characteristic age $\tau$=P/2$\pdot$$\sim$4,700 yr is consistent with that
of Kes~73, thus strenghtening the association. The X--ray spectrum of \ee~ is 
very soft (power law photon index $\sim$3.4), and its luminosity has remained 
remarkably constant at a level of $\sim$3 10$^{35}$ erg s$^{-1}$ (for d=7 kpc)
during all the observations carried out up to now (Gotthelf, Vasisht \& Dotani 
1999).

On the basis of these observational properties, \ee~  can be included in the 
class of Anomalous X--ray Pulsars (AXPs, Mereghetti \& Stella 1995). These 
pulsars, while clearly not powered by the     rotational energy losses, are 
characterized by soft spectra, periods in a narrow range (6 to 12 s), secular 
spin down, absence of massive companions, no radio emission (see Mereghetti 
2000 for a review). Three of the six known AXPs are associated to supernova 
remnants (SNRs).

The AXPs are clearly very different from the more common accreting X--ray 
pulsars usually found in High Mass X--ray Binaries, and their nature is still 
uncertain. Several models (involving both binary systems and isolated objects) 
have been proposed, but up to now the only reasonably well established fact is 
that AXPs are yet another manifestation of neutron stars.

Here we report a study of the possible counterparts of \ee~ based on deep 
optical and infrared observations
of its small error region.

\section[]{Observations }
 
Several observations of the field of \ee~   were carried out  using  different
European Southern Observatory (ESO) telescopes at La Silla from 1993
to 1999. Table\,\ref{tab:uno} gives a log of the observations that are 
described in more detail in the following subsections.

\begin{figure*} 
\mbox{} 
\vspace{12.5cm} 
\includegraphics{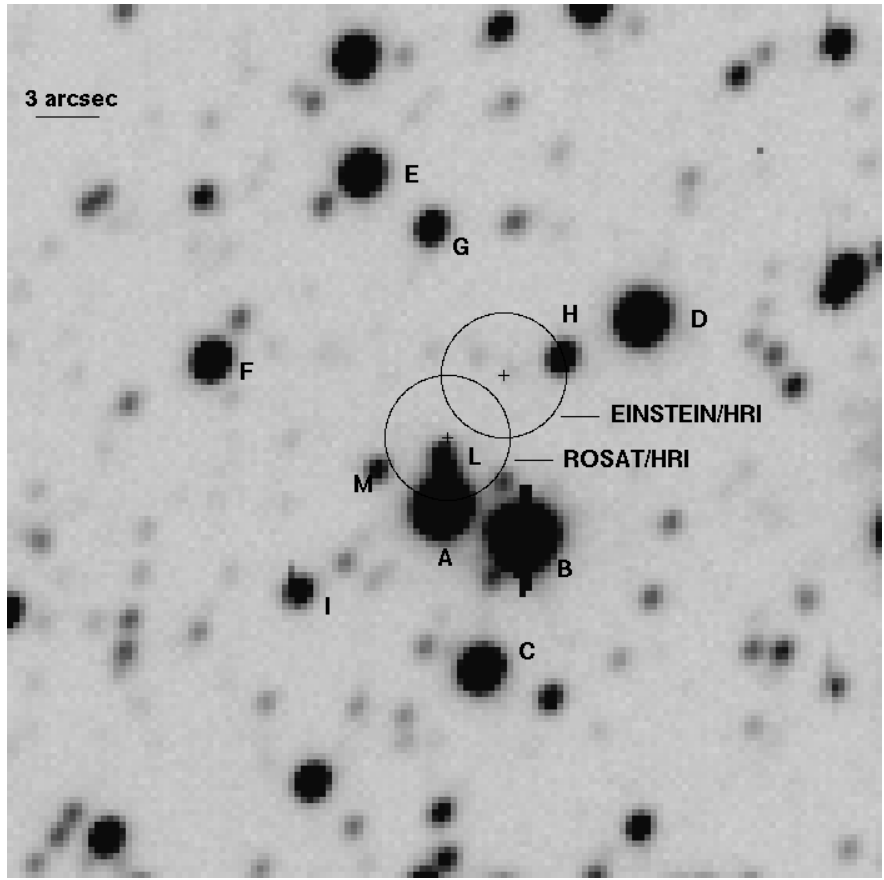} 
\caption[]{R band image of the field of \ee~ obtained with the NTT/EMMI in 
1993. Here and in the following images North is  top and East   left. 
The two error circles have radii of 3$''$.}
\label{fig:nttvla} 
\end{figure*}
 
\begin{figure*} 
\mbox{} 
\vspace{12.5cm} 
\includegraphics{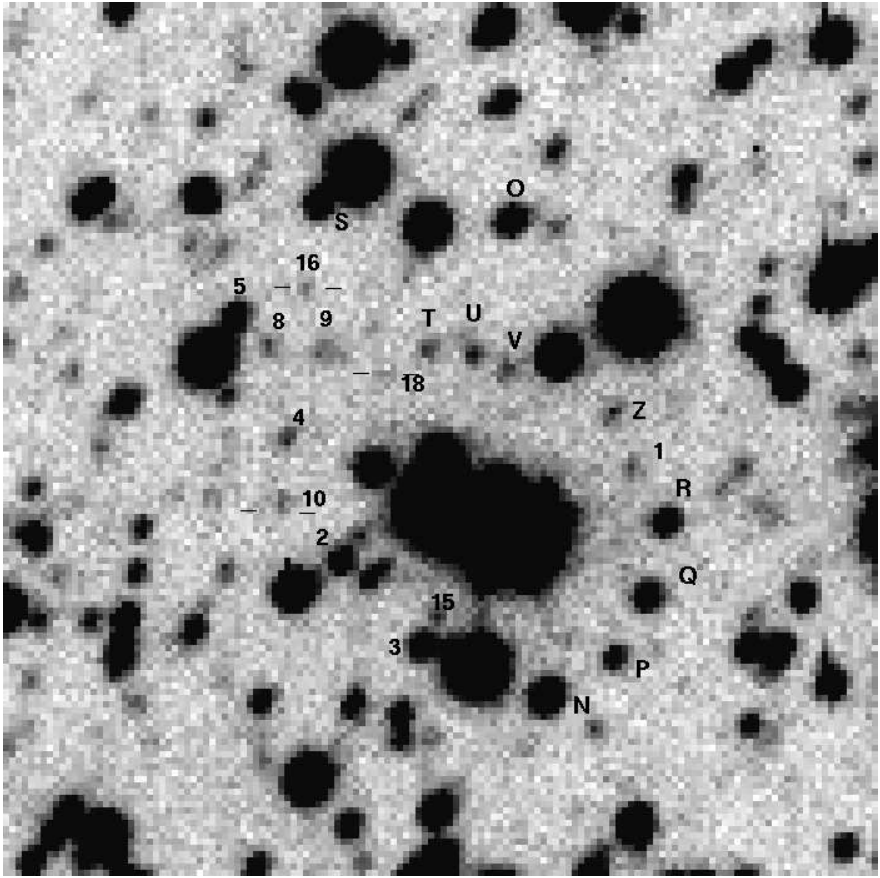} 
\caption[]{Deep R band image of the same field shown in Fig.\,\ref{fig:nttvla}.
The color scale has been stretched to show better the faintest stars.}
\label{fig:nttvlb} 
\end{figure*}

\begin{table}
\begin{center}
\begin{tabular}{|l|l|l|c|c|c|} \hline
{\em Date} & {\em Tel.} & {\em Instrument} & {\em Filter} & {\em Exp.} & 
{\em Seeing} \\
\hline
24.06.1993  & NTT  & EMMI   & V    & 10~min     & 1\farcs2    \\
24.06.1993  & NTT  & EMMI   & R    & 10~min     & 1\farcs0    \\
\hline
06.08.1997  & 2.2m & EFOSC2 & B    & 10~min     & 1\farcs5  \\
\hline
14.09.1999  & 3.6m & EFOSC2 & r    & 15~min     &  1\farcs1 \\ 
\hline
25.07.1999  & NTT  & SOFI   & J    &  1~min     & 0\farcs7 \\ 
25.07.1999  & NTT  & SOFI   & K$_s$&  1~min     & 0\farcs7  \\ 
\hline
12.04.2000  & OAN   &  La RUCA    & B   & 15~min     & 1\farcs5  \\ 
12.04.2000  & OAN   &  La RUCA    & V   &  5~min     & 1\farcs5  \\ 
12.04.2000  & OAN   &  La RUCA    & R   &  6~min     & 1\farcs5  \\ 
\hline
\end{tabular}
\vspace{1.0cm}
\caption{Summary of the imaging observations of the \ee~
field. Columns list the observing epoch,   telescope,  
imaging instrument,   filter,   total exposure time 
and   average seeing conditions during each observation.}
\label{tab:uno}
\end{center}
\end{table}

\subsection{Optical Imaging}

Images in the V and R filters were obtained with the 3.5~m NTT in June 1993, 
using the first generation of the ESO Multi Mode Instrument (EMMI) operated in
imaging mode with the red-sensitive CCD. 
The detector, a 2048$\times$2048 Loral CCD, had a field of view (f.o.v.) of
$10\arcmin \times 10\arcmin$ with a pixel size of 0\farcs35.  

The field of \ee~ was observed in the Bessel B filter using EFOSC2 at the 
ESO/MPG 2.2~m in August 1997. We used the 2048$\times$2048 Loral chip (ESO 
CCD No. 40) with a pixel scale of 0\farcs262 per pixel.

Finally, an image in the Gunn r filter was taken with the EFOSC2 instrument on 
the 3.6~m telescope in September 1999. We used the  2048$\times$2048 Loral/Lesser
(ESO CCD No. 40) with a  2$\times$2 rebinning which resulted in a pixel scale of 0\farcs31
per pixel.

All the data were reduced using standard procedures for bias subtraction, 
flat--field correction and cosmic rays cleaning. Profile fitting photometry was
carried out with the DAOPHOT\,II package (Stetson 1987) as implemented in 
ESO--MIDAS 98NOV version. Calibration was performed by standard stars observed,
as far as possible, in the same condition of the scientific frame. 
While the relative magnitude values from these data 
are quite accurate ($\sim$0.01--0.02 mag), it was not possible to derive
a precise absolute calibration since most of the  observations
were carried out in non-photometric conditions. 
Therefore, in order to obtain the absolute calibration, we reobserved
the field in the B, V and R filters under photometric conditions 
using the 1.5m telescope of the Observatorio Astron\'omico Nacional at San Pedro 
M\'artir (Baja California, Mexico) equipped with a $1024 \times 1024$ CCD 
detector with a pixel scale of 0\farcs24 per pixel. 
This allowed to obtain the results summarized in Table\,\ref{tab:due}.
The Gunn r photometry was calibrated by means of standard star observed in the
R band and the average relations between {\it uvgr} and UBVR systems (Kent 
1985).

\subsection{Infrared Imaging}

Near--infrared observations have been carried out at the NTT using SOFI. 
The SOFI instrument is an infrared spectrograph and imaging camera with a 
HgCdTe $1024 \times 1024$ array and a pixel size of 18.5 $\mu m$. The camera 
was operated with a f.o.v. of $4\farcm94 \times 4\farcm94$ and a corresponding 
pixel scale of  0\farcs292. Images have been taken through the wide-band J
($\lambda = 1.247 \mu m$; $\Delta \lambda = 0.290 \mu m$) and K$_{\rm s}$ 
($\lambda = 2.162 \mu m$; $\Delta \lambda = 0.275 \mu m$) filters with 
exposure times of 60 s. The typical seeing for these observations was 
$\sim$0.7$''$. After taking each image of the object, an image of the sky was
acquired, to permit the subtraction of the blank sky. The images were further 
treated by removal of the dark current,   flat field and   bright infrared 
sky.

\subsection{Optical spectroscopy}

Spectroscopy of the brightest candidate counterparts (objects A and B)
was obtained at the 1.5~m ESO Spectrographic telescope on July 1998, using a 
B\&C Spectrograph.  The exposure time was 20\,min;  the slit width was 2$''$.
The spectra covered the wavelength range 4000--8000\,\AA\ with a resolution 
of $\sim 10$\,\AA\ (FWHM). Spectra were flux calibrated under non-photometric 
conditions, so that just the relative intensities are   reliable. 

\begin{figure}  
\mbox{} 
\vspace{7.5cm} 
\includegraphics{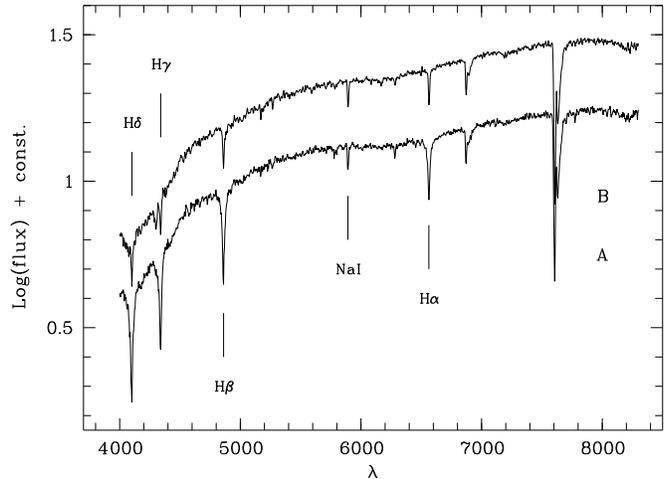} 
\caption[]{Flux calibrated spectra for candidates A (lower line) and B taken 
at the 1.5m Danish telescope.}
\label{fig:eso} 
\end{figure}

\begin{figure*}  
\mbox{} 
\vspace{10.5cm} 
\includegraphics{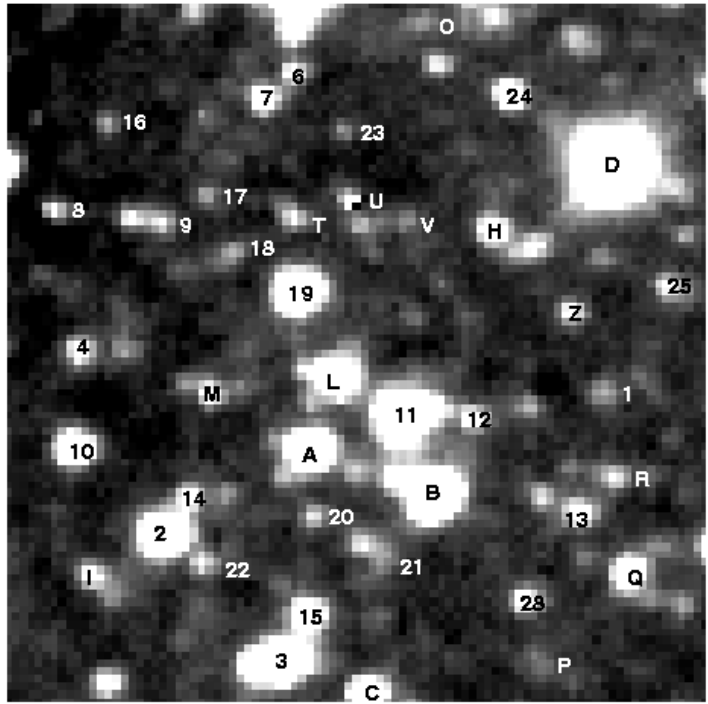} 
\caption[]{K$_s$ band image of the field of \ee~. }
\label{fig:irk} 
\end{figure*}

\begin{figure*} 
\mbox{} 
\vspace{10.0cm} 
\includegraphics{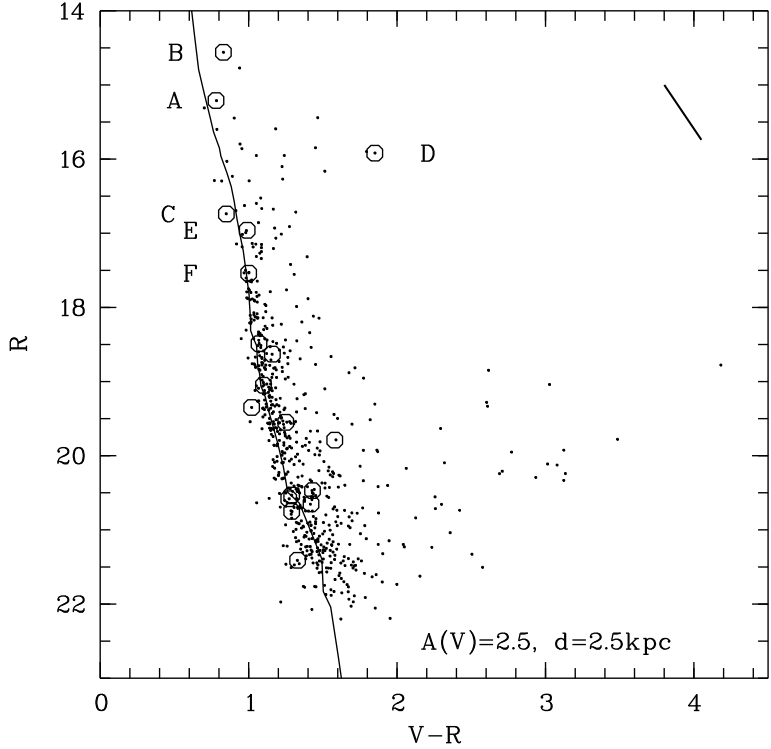} 
\caption[]{Optical color-magnitude diagram for the stars in a $\sim$3$'\times$3$'$
field around the position of \ee~. The circles mark the stars 
within or close to the error circles. Note that the stars brighter than 
R$\sim 18$ are partly saturated and their profile--fitted magnitudes are to be 
considered cautiosly.  A Pop.\,I MS at   2.5\,kpc and with a reddening 
$A_V \sim 2.5$ gives a good fit of the data, in agreement with the spectroscopic estimate 
for star A.
The segment in the upper right corner indicates the direction of reddening 
(the length corresponds to A$_v$ = 1 mag).}
\label{fig:vr} 
\end{figure*}
   
\begin{figure*} 
\mbox{} 
\vspace{9cm} 
\includegraphics{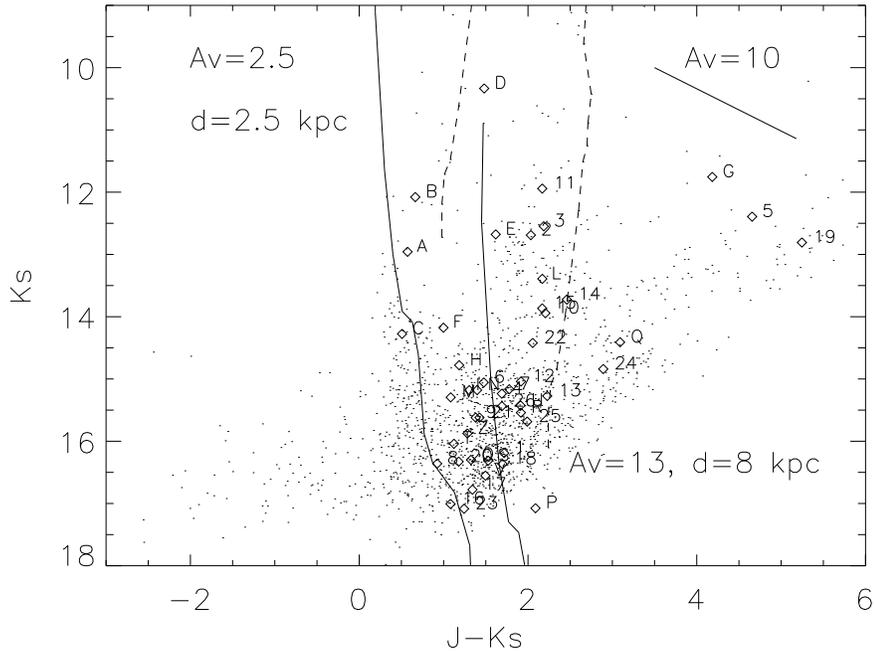} 
\caption[]{IR color magnitude diagram for the stars in the region of \ee~. 
The expected positions for main sequence stars (full lines) and red giants
(dashed lines) for selected values of the distance and reddening are also 
indicated. The diamonds mark objects close to or within the \ee~ error boxes.
The segment in the upper right cornes indicates the effect of reddening due to A$_v$ = 10 mag. }
\label{fig:jk} 
\end{figure*}

\begin{table}
\begin{center}
\begin{tabular}{|c|c|c|l|c|c|c|} \hline
 star &  B   &   V   &    R  & r Gunn &  J   &   K  \\
\hline
  A  & 17.14 & 15.95 & 15.21   & 16.21 & 13.53 & 12.96\\
  B  & 16.71 & 15.39 & 14.56   &       & 12.74 & 12.08\\
  C  & 18.93 & 17.55 & 16.74   & 17.32 & 14.78 & 14.28\\
  D  & 20.64 & 17.83 & 15.92   & 16.65 & 11.81 & 10.33\\
  E  & 19.49 & 17.96 & 16.96   & 17.66 & 14.30 & 12.68\\
  F  & 20.22 & 18.55 & 17.54   & 18.21 & 15.17 & 14.18\\
  G  & 21.39 & 19.57 & 18.49   & 19.21 & 15.94 & 11.76\\
  H  & 21.66 & 19.80 & 18.63   & 19.40 & 15.96 & 14.78\\
  I  & 21.91 & 20.16 & 19.05   & 19.57 & 16.48 & 15.18\\
  L  & 22.27 & 20.38 & 19.35   & 19.42 & 15.56 & 13.39\\
  M  & 23.51 & 21.43 & 19.79   & 20.34 & 16.38 & 15.30\\
  N  & 22.83 & 20.86 & 19.55   & 20.10 & 16.58 & 15.18\\
  O  &       & 21.88 & 20.53   & 21.07 & 17.62 & 16.30\\
  P  &       & 22.80 & 21.41   & 21.84 & 19.17 & 17.08\\
  Q  &       & 21.91 & 20.58   & 21.18 & 17.50 & 14.41\\
  R  &       & 22.10 & 20.75   & 21.18 & 17.46 & 15.54\\
  S  &       & 21.96 & 20.47   & 20.95 & 17.84 & 16.31\\
  T  &       &       & 22.37   & 23.02 & 17.16 & 16.04\\
  U  &       &       & 22.07   & 22.81 & 17.34 & 15.43\\
  V  &       &       & 22.32   & 22.94 & 18.05 & 16.55\\
  Z  &       &       & 22.34   &       & 17.16 & 15.88\\
  1  &       &       & 22.59   & 23.17 & 17.91 & 16.19\\
  2  &       &       & 20.84   & 21.51 & 14.72 & 12.69\\
  3  &       &       & 20.77   & 21.61 & 14.73 & 12.55\\
  4  &       &       & 22.27   & 23.14 & 16.93 & 15.24\\
  5  &       & 22.13 & 20.65   & 21.20 & 17.05 & 12.39\\
  6  &       &       &         &       & 16.53 & 15.06\\
  7  &       &       &         &       & 16.94 & 15.16\\
  8  &       &       & 22.52   & 23.98 & 17.28 & 16.36\\
  9  &       &       & 22.50   & 23.26 & 17.00 & 15.62\\
  10 &       &       &         &       & 16.15 & 13.94\\
  11 &       &       &         &       & 14.11 & 11.94\\
  12 &       &       &         &       & 16.96 & 15.04\\
  13 &       &       &         &       & 17.50 & 15.27\\
  14 &       &       &         &       & 16.18 & 13.73\\
  15 &       &       & 22.41   &       & 16.04 & 13.87\\
  16 &       &       & 23.45   &       & 18.09 & 17.01\\
  17 &       &       &         &       & 18.12 & 16.78\\
  18 &       &       & 23.89   &       & 18.07 & 16.37\\
  19 &       &       &         &       & 18.05 & 12.81\\
  20 &       &       &         &       & 17.51 & 16.33\\
  21 &       &       &         &       & 17.05 & 15.62\\
  22 &       &       &         &       & 16.48 & 14.42\\
  23 &       &       &         &       & 18.33 & 17.08\\
  24 &       &       &         &       & 17.73 & 14.84\\
  25 &       &       &         &       & 17.68 & 15.69\\
  26 &       &       &         &       & 17.13 & 15.44\\
  27 &       &       &         &       & 17.38 &      \\
  28 &       &       &         &       &       & 15.59\\  
\hline
 
\end{tabular}
\caption{Magnitudes of the stars close to the position of \ee~.}
\label{tab:due} 
\end{center}
\end{table}

\section{Results}

An R--band image of the field of \ee~ is shown in Fig\,\ref{fig:nttvla}. The two 
error circles are centered at the positions obtained with the ROSAT HRI, 
R.A.=18$^h$ 41$^m$ 19.2$^s$  Dec. = --04$^{\circ}$ 56$'$ 12.5$''$ (J2000, 
Helfand et al. 1994) and with the Einstein HRI, R.A.=18$^h$ 41$^m$ 19.0$^s$  
Dec. = --04$^{\circ}$ 56$'$ 08.9$''$ (J2000, Kriss et al. 1985). Both positions
were reported with uncertainties of 3$\arcsec$, but possible errors in the 
satellites boresight cannot be excluded.
For this reason we considered  all the objects
within $\sim$10$''$ from the nominal positions.
Our astrometry, computed with the ASTROM software (Wallace 1990) 
on a short exposure EMMI image by using as a reference a set of stars from the 
DSS2, 
has an overall accuracy of $\sim$1$\arcsec$.

Several objects are present within or close to the error circles. Their 
magnitudes are reported in Table\,\ref{tab:due}. A deeper image is shown in 
Fig.\,\ref{fig:nttvlb}. 

The spectra for the brightest objects, B and A, are shown in 
Fig.\,\ref{fig:eso}. Classification was performed by comparison with the
spectral library of Silva \& Cornell (1992). Based on this 
we classify star B as a F6--7\,III and star A as A5--8\,V. Both spectra are 
quite reddened. From the EW of the NaI lines, we can estimate an interstellar 
absorption larger than A$_{V}\sim$1.5 for both stars (Munari \& Zwitter 
1997). 

The infrared image in the K$_s$ band is shown in   Fig.\,\ref{fig:irk}.
Tanks to the less severe
absorption at long wavelengths, these data give the deeper view of the
region, revealing several objects undetected in the visible band.

\section{Discussion}

The distance of Kes~73 was determined by means of the 21~cm absorption
measurements by Sanbonmatsu \& Helfand (1992), who obtained a value
of 6--7.5 kpc. These authors also estimated an HI column density to the
source of $\sim$5 10$^{21}$ cm$^{-2}$. The equivalent hydrogen column density 
obtained from X--ray spectral fits of \ee~ and Kes~73 is in the   
$\sim$1.5-3 10$^{22}$ cm$^{-2}$ range (Gotthelf \& Vasisht 1997). It is thus clear 
that we can expect a significant reddening for the optical counterpart of \ee~.
For instance assuming N$_H$=2 10$^{22}$ cm$^{-2}$ and using the average 
relation N$_H$ = 1.79 10$^{22}$A$_V$ cm$^{-2}$ (Predehl \& Schmitt, 1995)
we expect A$_{V}\sim$11 (note however that there is a large scatter 
around the above average relation).

The two brightest objects we studied (A and B) are normal stars  without any
distinctive feature in their spectra that might suggest an association with 
an X--ray pulsar. Their magnitudes and colors also indicate a relatively 
small distance of $\sim$3 kpc.

The color-magnitude diagram based on the V and R NTT observations is shown 
in Fig.\,\ref{fig:vr}. The circles indicate the objects within or close to 
the error regions; their magnitudes  are given in Table\,\ref{tab:due}.
None of these objects possesses especially    unusual colors. The V-R color--magnitude 
diagram, that shows a clearly defined sequence, is very similar to that of a 
star cluster, indicating that most of the stars that we   observed have 
similar distance and reddening.

The possible presence of a star cluster would be particularly interesting;
however, a more likely explanation is that we are looking at stars belonging
to one of the spiral arms of our Galaxy. In fact,  the line of sight toward 
Kes~73 intercepts  Arm 3 at  $\sim$2.5-3 kpc (see, e.g., Taylor \& Cordes 1993)
and then is almost tangential to Arm 2 for distances between $\sim$5 and 
$\sim$8 kpc. For comparison, the lines corresponding to a main sequence for
different values of reddening and distance are also plotted in 
Fig.\,\ref{fig:vr}. It is clear that most of the points in the diagram 
correspond to stars in the spiral arm at $\sim$2.5-3 kpc, while only a few 
farther objects are visible due to the strong reddening in the Galactic plane.
Therefore, if we trust the association between \ee~ and the SNR at d$\sim$7 kpc,
most of the objects seen in the optical images are likely foreground stars 
much closer than Kes~73.

The presence of a large number of stars with similar values of distance and reddening
is also evident from the IR color-magnitude plot shown in Fig.\,\ref{fig:jk}. 
Here  objects consistent with a red giant branch are also visible, as well as 
many other more distant and/or reddened stars that are not seen in the optical 
bands. The objects within or close to the error boxes of \ee~ are indicated
with diamonds. 
Three of them are very reddened (G, 5 and 19), with colors not compatible with
main sequence stars. They are probably very distant giant or supergiant stars. 
In particular, while stars G and 5 are quite distant from the error regions of 
\ee~, star 19 should be considered as a potentially interesting candidate counterpart.

Many accreting pulsars have supergiant OB companions. However, the IR 
magnitudes of object 19 does not support an early type star. 
OB supergiants have typically (J--K) in the --0.2 to  0.1 range (Johnson 1966, 
Ruelas-Mayorga 1991) therefore a very high absorption (A$_v$ $\gsim$25-30) is 
required by the observed value of (J--K)=5.24$\pm$0.12, as well as a distance placing it 
at the edge of the Galaxy, well beyond Kes~73. The possibility of a Be system 
seems also unlikely: though a smaller distance is allowed (e.g. $\sim$10 kpc 
for a B5) still a very large A$_V$ would be required,
(of course, a smaller  A$_V$  could be assumed in the presence of 
substantial IR emission from  circumstellar dust). A more plausible 
explanation for star 19 is that of a very reddened giant at d$\gsim$8 kpc.

Some models for  AXPs involve isolated neutron stars accreting
from residual disks (van Paradijs et al. 1995, Ghosh et al. 1997, Chatterjee et al. 1999). 
Although it is unclear whether such disks can actually be formed, (e.g. in the 
evolution of Thorne--Zytkow objects or as a result of material fallback after 
a supernova explosion), and whether they can explain all the observed 
properties of the AXP, Perna et al. (2000) made some predictions on the 
emitted optical and IR spectra. Very different optical and IR colors can be 
obtained by the models presented by these authors, as a function, e.g., of the 
neutron star age and magnetic field, inclination and mass of the disk, 
importance of reirradiation of the pulsar X--ray flux, etc. Therefore
it is not surprising that the 
current data  do not support nor weaken such models. For instance, the values 
K$\sim$17 and V$\sim$28 predicted by Perna et al. for \ee~ assuming an age of 
$\sim$4000 yrs, B = 8 10$^{12}$ G, a disk mass of 0.005 $\msole$ and an 
inclination of 60$\deg$ are consistent with several of the objects visible in 
our IR data and below the detection threshold in the optical.

\section{Conclusions}

The optical limits on the possible counterparts of other AXPs have allowed to 
rule out the presence of a (non-degenerate) massive companion (see Mereghetti 
2000 and references therein). Unfortunately, we cannot do the same yet for \ee,
based on the observations presented here.
However, an OB supergiant companion seems unlikely, as it  requires values of the 
distance and/or absorption much in excess of those estimated for \ee~ to be 
compatible with the objects we found.
Massive main sequence stars cannot be excluded. For instance a B5 star at 
$\sim$7 kpc and with A$_{v}\sim$8 would have J and K magnitudes fully 
consistent with e.g., those of star H.
Although the optical and IR data cannot exclude the possibility that \ee~ be 
a Be binary, its X--ray properties are very different from those of such a class
of sources, that are characterized by hard X--ray spectra, significant variability
and erratic long term evolution in the   spin  period.  

None of the objects we studied has peculiar characteristics that might suggest
a possible connection with \ee~. However, if this source is indeed associated 
with Kes~73 and therefore subject to a significant reddening, we have 
discovered within its error region a promising candidate (star 19) deserving 
further spectroscopic study. 
We expect that the data 
presented here will represent a useful reference to 
facilitate the identification task, once the positional accuracy
of  \ee~  will be much improved by observations
with the new generation X--ray satellites $Chandra$ and 
$NewtonXMM$.
  
\section*{Acknowledgments}
S.Ch.  acknowledges financial support from grant  F/00-180/A 
from the Leverhulme Trust. We 
acknowledge  the support software  provided by the  Starlink Project which is 
funded by the UK SERC.

\label{lastpage}

\end{document}